# Expected vibroacoustic behaviour of Greek Doric-style temples and its relation with geometrical physics design as part of the intangible cultural heritage

Fabrizio Barone [1, *], Marco Casazza [1]

[1]Department of Medicine, Surgery and Dentistry "Scuola Medica Salernitana", Baronissi, Italy

Corresponding author: fbarone@unisa.it

**Abstract.**

This study proposes a new approach to the interpretation of Greek Doric-style temples, based on the integration of its tangible and intangible dimensions as a cultural heritage asset. Rooted on the Greek concept of techne (η τέχνη), the work considers a unifying design principle, integrating both structural and functional aspects within the architectural style. A multidisciplinary perspective was adopted, combining archaeological, documentary, and metrological analysis of 41 Doric temples from the 6th to the 4th century BC, located in Greece and Magna Graecia. Starting from the evidence of a statistical correlation among key geometric parameters, these quantitative data are re-interpreted through a geometrical physics vibro-acoustic model. The results demonstrate that structural elements act as acoustic attenuators, minimizing environmental forces — particularly wind — on the temple's cell walls. The study also suggests that slight deviations from the classic East-West orientation were adopted to reduce the acoustic coupling with prevailing local winds. The Archaeological Park of Paestum (Salerno, Italy) provides significant evidence for this hypothesis, as its temples, despite their different construction periods, share a consistent orientation, distinct from the city's street grid. These findings contribute to a deeper understanding of Greek know-how, being a part of the intangible dimension of cultural heritage and traditional ecological knowledge related to the architectural design in relation to the environmental factors.

## 1. Introduction

The modern interpretation of cultural heritage as a synergy of material and immaterial dimensions have prompted a re-evaluation of archaeological and documentary findings through multidisciplinary inverse analysis techniques (Vecco, 2010; UNESCO, 2013). This approach enables a deeper understanding of historical trajectories that shaped cultural and environmental identities (*genius loci*), beyond conventional archaeological and historical interpretations.

The study, overcoming the fragmented modern vision of culture, aligned with the classical concept of techne (η τέχνη), a fusion of intellectual and manual expertise, aimed at achieving specific functional goals. This new multidimensional vision broadening also the understanding of material heritage remains, that cannot be confined within a sub-space defined by their aesthetic interpretation, based on archaeological evidence and historical documentation, while relegating the geometric and structural features to an ancillary role.

To investigate the interplay between the material and immaterial dimensions of cultural heritage, we analysed a sample of Greek Doric temples using a structured methodology. Rather than limiting Doric temples to their aesthetic and religious significance, this study interprets them as manifestations of techne, where geometric and structural choices reflect functional design principles. By re-analysing the temples geometric data, we moved from the existing correlations, justifying the hypothesis of a unified architectural design model. Then, we introduced and validate a vibroacoustic geometric model, demonstrating that Doric temple design had the effect of minimizing the effects of environmental forces on the structures, offering also new insights on their orientation.

## 2. Geometric analysis of Doric temples

The work started collecting a database of the geometric measures of 41 Doric temples built between the 6$^{th}$ to the 4$^{th}$ century BC and located in Greece and Southern Italy (the area historically defined as Magna Graecia). The data were freely available from the literature (Woodward, 2013). These data, re-ordered, are made available as Supplementary Materials. These data were used to investigate the geometric relationships among the main structural elements of the temples (form factors), neglecting the individual temples' architectural specificities (friezes, decorations, religious function, etc.). Metrology (geometric dimension) and statistics (linear regression based on the least-squares method) constituted the basis of this approach, avoiding the introduction of any aesthetical consideration. Specific correlations are expected, knowing, from the literature, that proportions were used by Greek architects (Pakkanen, 2013; Wilson Jones, 2015).

Figure1 shows the typical scheme of a Doric temple, that highlights its external shape main parameters: Figure 1(a) represents a typical temple horizontal section, while Figure 1(b) displays the simplified three-dimensional geometric model used for this study, consisting in a superposition of four "equivalent" parallelepipeds (base, colonnade, entablature) and a triangular prism (pediment), positioned one above the other. The analysis was limited to the base (red) and to the colonnade (light blue) parallelepipeds to confirm the existence of temples' geometric similarities, regardless of their location and relevance, and to build an effective functional vibroacoustic model of a Doric temple.

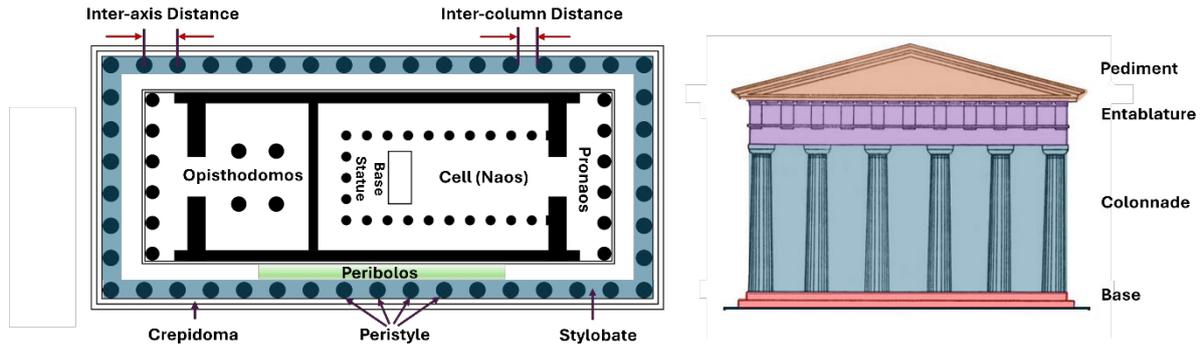

**Figure 1.** Typical geometric scheme of a Greek doric temple: a) horizontal section (left), b) simplified three-dimensional geometric model (right).

The results of the analysis proved that the temple components (column height, distance between columns, height of temple, etc.) are multiples or submultiples of a module (base diameter of the column), used as unit of measurement. Figure 2 shows the best fit curves correlating the temples' bases, stylobates and cells dimensions. The Temple of Zeus at Akragas (Agrigento, Italy), available in the database was excluded from the statistical analysis due to its different architecture.

These relationships can be analytically described by the relations:

$$b_w = 3.64 \cdot 10^{-1} \cdot b_l + 3.86 \; m \qquad (1)$$

$$s_w = 3.42 \cdot 10^{-1} \cdot s_l + 3.71 \; m \qquad (2)$$

$$c_w = 2.50 \cdot 10^{-3} \cdot c_l + 3.05 \; m \qquad (3)$$

where $b_w, s_w, c_w$ and $b_l, s_l, c_l$ are widths and lengths of the bases, stylobates and cells, respectively.

The base structural correlations served to guarantee the structural stability in face of seismic forcing. The observed deviations from a linear correlation between the base dimensions the geometric dimensions of the base parallelepiped bases might depend on the technical solution used for the temple foundation. Instead, the cell dimensions variability might vary with the functional religious and aesthetic requirements.

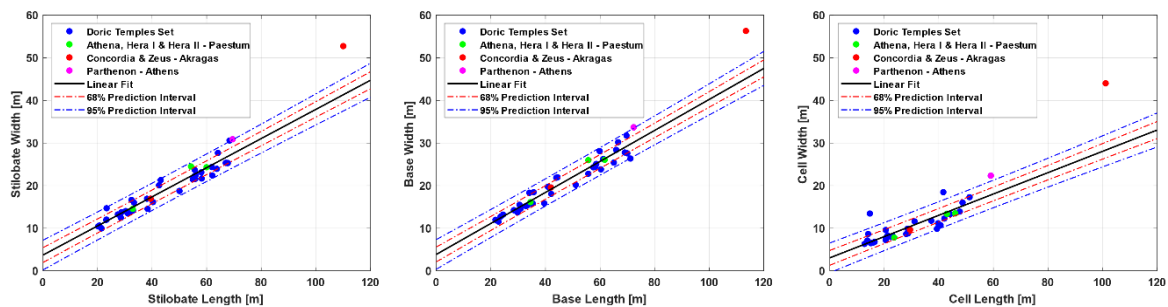

**Figure 2.** Linear regression curves (continuous black) showing the correlations between the horizontal lengths and widths for the Doric temples parameters: base (top), stylobate (centre) and cells (bottom). Two other relevant correlation curves are also shown: dashed red curve, delimiting the 68% level of confidence; dashed blue curve, delimiting the 95% level of confidence.

The colonnade (light blue in Figure 1b), including the lateral and front sides dimensions, is geometrically defined by column number, height and diameters (top and bottom) as well as by the intercolumn distance, all relevant geometric for the physical and aesthetic structure of a temple. Figure 3 shows the linear correlations between column heights and stylobate lengths (Figure 3(a)), the column base (stylobate) and heights (Figure 3(b)) and the column base (stylobate) and top (architrave) diameters (Figure 3(c)) for the temples' lateral side.

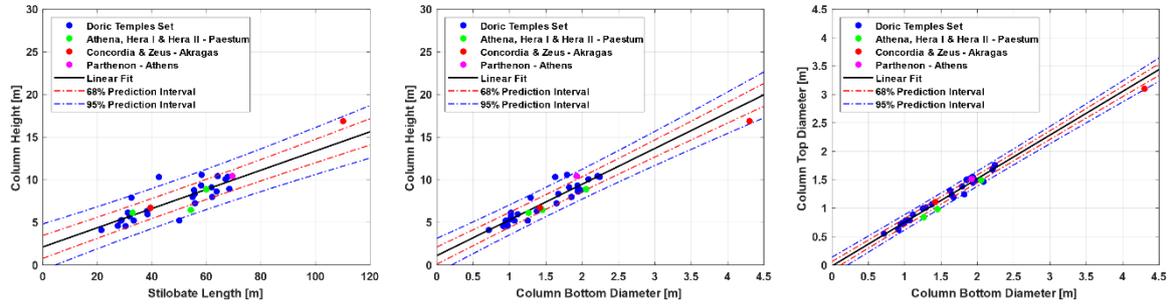

**Figure 3.** Linear regression curves (continuous black) showing the correlations between the stylobate length and the column height (left), the column bottom diameter and column height (centre), the column bottom and top diameters (right) for the Doric temples set. Two other relevant correlation curves are also shown: dashed red curve, delimiting the 68% level of confidence; dashed blue curve, delimiting the 95% level of confidence.

These relationships can be analytically described by

$$c_H = 1.12 \cdot 10^{-1} \cdot s_l + 2.13 \; m \tag{4}$$

$$c_H = 4.19 \cdot c_{BD} + 1.11 \; m \tag{5}$$

$$c_{TD} = 7.69 \cdot 10^{-1} \cdot s_l - 1.93 \cdot 10^{-2} \; m \tag{6}$$

where $c_H, c_{TD}, c_{BD}$ are the column top diameter, bottom diameter and height, respectively. Similar correlations were found also for the frontal colonnade sides.

These results lead to a first preliminary conclusion that the construction of a temple can always be traced back to a simple basic scheme, at least as far as the structural and architectural parameters are concerned. On the other hand, upon observation of the internal frontal (front and back) temple architecture, large differences in construction emerge, which could be attributable to architectural, aesthetic and religious reasons. These differences are already statistically evident, expressed by deviations from the linearity of the statistical curves related to the main parameters analysed. Otherwise, could these differences be indicators of objectives that can be presently only hypothesized, achieved through undocumented technology, part of an intangible heritage to be recovered?

# 3. Vibroacoustic interpretation of a Doric Temple

## 3.1 Introduction

In the Greek vision of techne, the artistic-architectural and technical aspects of a temple are inseparable. Its design reflects multiple converging objectives: aesthetic and architectural beauty as expressions of culture and wealth; durability and resistance to environmental forces as markers of technological advancement; and the acoustic isolation of the cell (in Greek, *o ναός*) as a sacred space, simultaneously secluded and accessible.

While traditional interpretations focus on aesthetics and durability, this study extends the analysis to a vibroacoustic perspective, hypothesizing that Doric temples were designed to insulate the cell from external vibrations. Such isolation could only be achieved through a seismo-acoustic solution, decoupling the internal space from environmental forces. These forces, including wind and sea waves, generate broadband vibroacoustic signals composed of structural signatures and forced vibrations (Barone and Casazza, 2024).

Previous research demonstrated that Doric temples incorporated architectural solutions to minimize environmental forcing, whether acoustic (wind) or seismic (ground motion). Building on statistical metrological analysis, this study employs geometrical physics modelling, a field well understood by the Greeks (e.g., geometric optics and acoustics), to show that temple architecture, positioning, and orientation were functionally designed to mitigate external forces (Darrigol, 2010).

## 3.2 Research hypotheses and questions

Based on the visible parts of a structure (above the ground level), the statistical results highlight a unitary architectural geometric design of Doric temples. Despite this fact, coherent with a geometrical physics interpretation of the adopted vibroacoustic hypothesis, the possibility of identifying a solution, aimed at implementing vibro-acoustical decoupling (seismic and acoustic) of the cell inner space from the outside environment, remained unexplored.

This hypothesis is initially supported by the Greek deep knowledge of geometric acoustics, and its application, as techne, to the construction of theatres.

The second observation relates to the structure of the Greek temple basements, that behaves as a high-performance dynamic seismic attenuator, highlighting unexpected but very deep skills of the Greeks in the design of seismic (vibration) dampers. This is confirmed by the excavations of two temples in the Archaeological Park of Paestum and Velia. The damper typically consists of three layers: a natural layer (the natural travertine base, horizontally levelled), a sand layer (distributed and compacted on the travertine natural layer) and the temple's basement (a 3 m high parallelepiped built with limestone blocks) (Klein, 2016; Petti et al., 2019). A similar structure is currently used for state-of-the-art very low frequency seismic dampers for optical benches hosting stabilized lasers (Blom et al., 2015). The seismic attenuator certainly contributed to the vibration insulation at the bottom of the cell, reducing the amplitude of secondary acoustic modes of its inner part, that are observed in the case of vibratory excitations of a structure (Chang et al., 2009).

Moreover, Greeks practical wisdom on sailing, their knowledge on the Mediterranean Sea and its winds (periodicity, intensity, direction) as well as on astronomy and, locally, with the orography, could have played a role in temples' positioning. The deviations from a traditional East-West orientation, with maximum angular difference from east-west direction up to about 15°, cannot be attributed to a lack of orientation ability, considering the practical wisdom of

Greeks when the temples were built. Consequently, a different hypothesis, consisting in a precise design indication, should be explored.

These considerations lead to formulate two main science questions: Why are Doric temples not perfectly east-west oriented, but only apparently casually partly misaligned?"; Were the cell external walls insulated from the environment also acoustically? If yes, in which way?".

### 3.3 Premise to the acoustic module of the vibroacoustic model

The hypothesis that the architectural-structural elements of a Doric temple were functionally designed to be also acoustically insulated requires the adoption of a unifying vibroacoustic model derived from geometric considerations, considering that the geometric approach to acoustics, as well as optics, was well-known by Greeks and that a geometric approach is still physically valid (Gibbons and Warnick, 2011). In particular, the physical representation of acoustic pressure waves propagation and wind can be approached geometrically. Within this hypothesis, wind constitutes the strongest pressure forcing acting on a temple, capable of stressing front and lateral vertical walls also at very low frequencies.

The possibility of adopting the same approach to represent seismic and acoustic forcing on the temple cell, allows us to treat both the seismic and the hypothesized acoustic attenuation in the same way, considering the external colonnade as the attenuator main element.

This hypothesis led to the development of a geometric model, necessary for framing the temple in the environmental reality, assuming winds as the strongest acoustic forcing (pressure forcing).

The action of these forces on the front and lateral surfaces of the base (parallelogram), for the entablature (parallelogram) and for the pediment (triangular prism), can be evaluated by multiplying the areas of the exposed surfaces to the acoustic (wind) pressure. However, the real pressure exerted by winds on the cell walls is only part of the total pressure forcing, since it is partly intercepted by the external colonnade, reducing the cell wall surfaces directly exposed to winds. Hence, the geometric acoustic model needs to consider the wind incidence angle on the structure. The following sub-sections will detail the models of the temple front and lateral sides.

**3.4 Lateral side acoustic sub-model**

Geometrically, the two lateral surfaces can be modelled as an alternate sequence of closed and open surfaces, whose areas are defined by the sections of the columns parallel to the ideal lateral plane of the equivalent parallelepiped, as shown in Figure 1, and by empty surfaces, whose areas are defined by the free space between adjacent columns.

Figure 4 shows the change of the lateral geometric view of the cell for an observer, if, moving from the lateral side of the temple, at -90° position (Position A - lateral vision), where the maximum cell surface is visible, to 0° (Position D), where the cell surface is fully shadowed by the colonnade, gradually moving towards visible surfaces of increasingly smaller size (e.g. -40° - Position B and -20° - Position C).

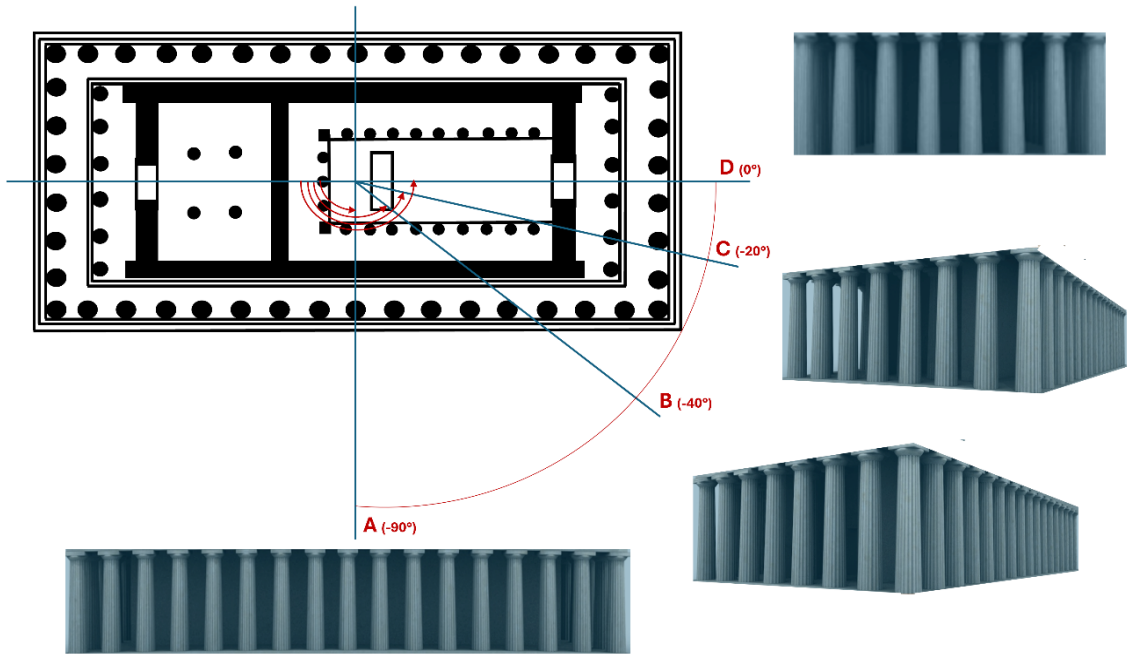

**Figure 4.** Map view of the winds action angle on a lateral side of the Athena Temple (Parthenon) in Athens, Greece for different observer's view angles: −90° (A), −40° (B), 20° (C), 0°(D). Note: the observer is considered positioned at a great distance to reproduce the typical "parallel forcing" of the wind source.

Figure 5a shows the geometric model section expressing the inter-column space, $l_{IC}$, as a function of the observer's angle of view (e.g. wind direction), $\theta$. As expected, the model shows that the inter-column space decreases at the decrease of the view angle. When the observer's axis of view is perpendicular to the plane identified by the lateral face of the temple colonnade ($\theta = 90°$), the inter-column space is maximum, decreasing as the observation angle decreases until it cancels at a critical angle, $\theta_C$. Below these values, the lateral colonnade appears to the observer as a continuous lateral surface, precluding the view of "its interior". Therefore, for view angles, below the critical one, $\theta_C$, the peribolos, that is the area surrounding the internal lateral sacred spaces of the temple, appears as a corridor limited by two walls, that are the cell wall and the "colonnade wall".

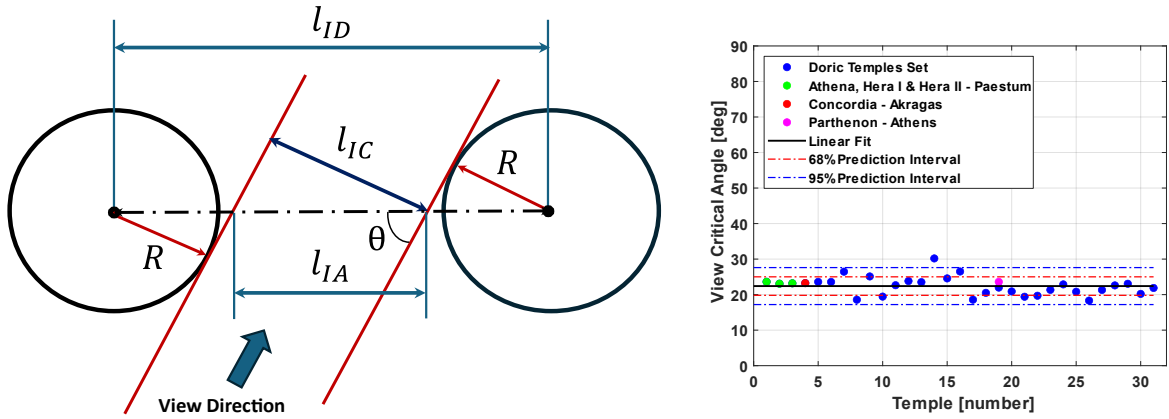

**Figure 5.** (a) Geometric model describing the intercolumniation distance as function of observer View Angle, $l_{IC} = l_{IC}(\theta)$ (left); (b) View Critical Angle, $\theta_c$, evaluated using the column mean diameter of a subset of 31 Doric temples (for which this information is available), together with the relative fit curves (continuous black) and two relevant confidence curves: dashed red, delimiting the 68% level of confidence, and dashed blue, delimiting the 95% level of confidence (right).

A relevant consequence of this geometry is that the interior of the cell is protected on this side by an internal wall, the cell wall, and by a 'virtual' external wall, the colonnade, separated by an air attenuator, the peribolos, that is the typical structure of an acoustic attenuator. Analytically, the inter-column space can be expressed as:

$$l_{IC} = l_{IC}(\theta) = l_{IA} \cdot \sin\theta = \left(l_{ID} - \frac{d}{\sin\theta}\right) \cdot \sin\theta = l_{ID} \cdot \sin\theta - d \qquad [\theta > \theta_c] \qquad (5)$$

$$\theta_{L_c} = \sin^{-1}\left(\frac{d}{l_{ID}}\right) \qquad (6)$$

where $l_{ID}$ is the linear distance between the geometric centers of adjacent columns, $d$ it is the mean diameter of the columns of the temple.

A strong relation exists also between the geometry of Doric temples and their critical angles, $\theta_{L_c}$. In fact, critical angles span from $18.32°$ to $30.19°$, with a distribution synthesized by mean and standard deviation, $\theta_{L_c} = 22.20° \pm 2.52°$ (Figure 5b).

The total wind impact surfaces (cross sections), are shown in Figure 6, both for the colonnade (Figure 6a), $S_{col_l}(\theta)$, and for the inter-column side (Figure 6b), $S_{inter_l}(\theta)$, as function of the view angle (wind direction), $\theta$, evaluated according to the relations

$$S_{col_l}(\theta) = \begin{cases} N_{LC} \cdot d \cdot \sin\theta & [\theta \leq \theta_C] \\ N_{LC} \cdot d & [\theta > \theta_C] \end{cases} \quad (7)$$

$$S_{inter_l}(\theta) = l_{IC}(\theta) \cdot c_H \cdot (N_{LC} - 1) \quad (8)$$

where $N_{LC}$ is the number of lateral columns.

Therefore, the column cross section increases with the angle, $\theta$, reaching its maximum value at ($\theta = \theta_{L_C}$), constant up to ($\theta = 90°$) due to their longitudinal symmetry, while the inter-column cross section starts to increase for $\theta > \theta_{L_C}$, reaching its maximum at $\theta = 90°$.

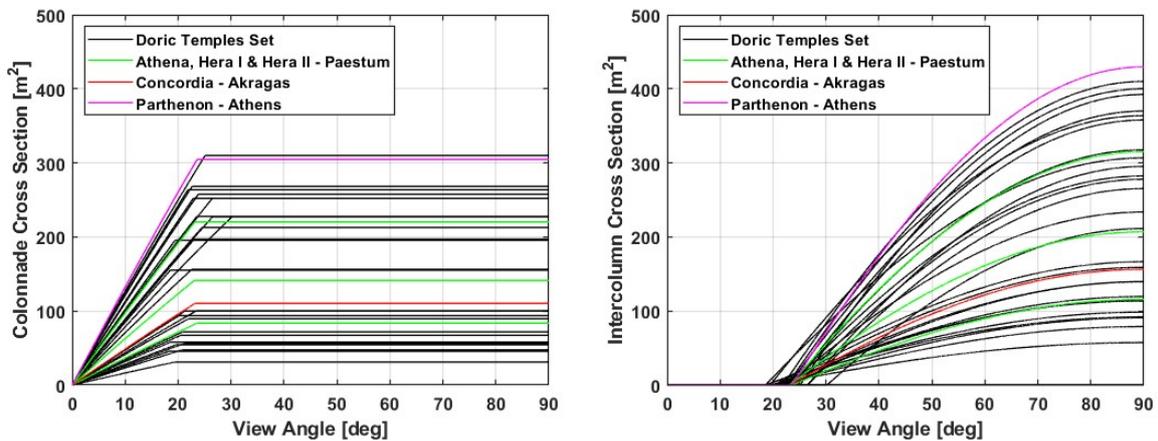

**Figure 6.** Geometric relations among lateral column total surface vs. the angle of View (left) and among the lateral intercolumn surface vs. the View angle (right) for all the Doric temples of the set.

The effective lateral cell wall cross section, $S_{cell_l}(\theta)$ is actually lower than $S_{inter_l}(\theta)$, both because the side walls lengths of the cells are lower than the stylobate lengths and because the effective incoming wind forcing on the cell, $P_{in}$, is determined only by its perpendicular component, $P_\perp$, being the parallel one, $P_\parallel$, channeled along the peribolos, with no actions on the cell wall (Figure 7a). Therefore, $S_{cell_l}(\theta)$ can be expressed by the relation

$$S_{cell_l}(\theta) = S_{inter_l}(\theta) \cdot \left(\frac{c_l}{s_l}\right) \cdot \sin\theta. \tag{9}$$

The trend of the effective cross section, $S_{cell_l}(\theta)$, as a function of the wind incidence angle, $\theta$, is shown in Figure 7b, while its percentage increment, referred to the maximum value, $S_{cell_l}(90°)$, is shown in Figure 7c.

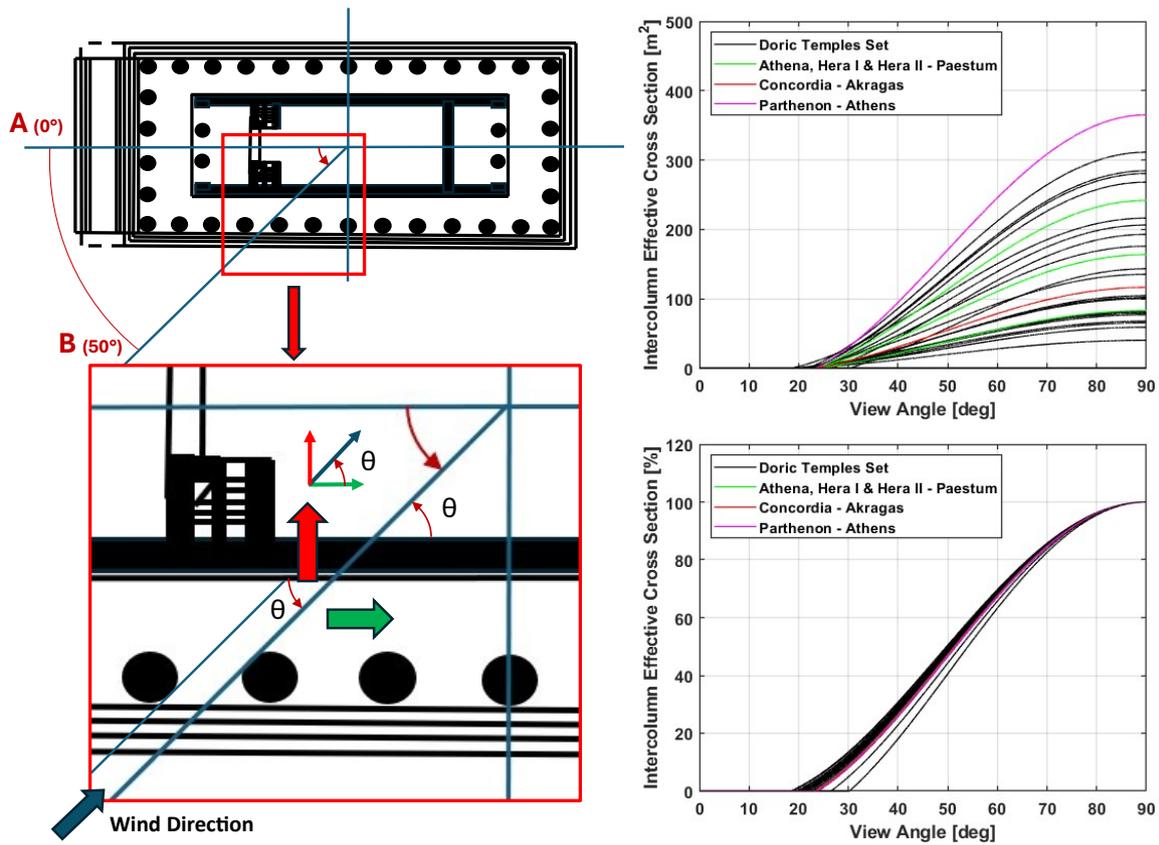

**Figure 7.** Concordia Temple: (a) geometric representation of the wind forcing effect on the external cell (left); (b) inter-column effective surface, $S_{cell_l}(\theta)$, as function of the wind incidence angle on the temple, $\theta$; (c) percentage of the effective cross-section, $S_{cell_l}(\theta)$, with respect to the maximum effective cross-section. $S_{cell_l}(90°)$, as function of the wind incidence angle on the temple, $\theta$.

Based on the model, the best acoustic insulation is obtained when the wind forcing pressure impinges the lateral side of the temple at an incidence angle between in the range [0°- $\theta_{Lc}$], that is when the lateral colonnade behaves as a closed continuous wall and, consequently, the peribolos acts as an equivalent closed corridor. Due to the symmetry of the structure, colonnade, peribolos and cell external wall behave as an acoustic attenuator for winds, whose direction are within the following two ranges [-$\theta_{Lc}$, $\theta_{Lc}$] and [180° - $\theta_{Lc}$, 180°+ $\theta_{Lc}$].

The effective lateral surface of the cell affected by wind pressure is very small up to viewing angles of the order of 30°. Figure 7c, in fact, shows that the percentage ratios between the effective intercolumn and maximum effective surfaces are less than 20% for viewing angles of the order of 35°, showing that the wind pressure exerted on the cell is still attenuated by 80%, indicating that acoustic decoupling can still be considered efficient even for relatively high $\theta$ angles.

### 3.5 Frontal and rear acoustic sub-model

The lateral colonnade geometric sub-model can be directly applied to the orthogonal frontal colonnade, naming the frontal incidence angle, $\theta'$. Nevertheless, this orthogonality allowed us to merge the sub-model into a single model. In fact, the angular transformation

$$\theta = 90° - \theta', \qquad (10)$$

allows to build a single model referred to the lateral colonnade reference system, defined by the two main axes of the temple (longitudinal and transverse), and initial angle ($\theta = 0°$), angle of incidence of the wind coming from the east-west direction, with positive counterclockwise rotation (Figure 5). Also in this case, the front colonnade behaves as an acoustic attenuator for wind directions within the ranges [90°- $\theta_{F_C}$, 90°], [-90°, -90° + $\theta_{F_C}$], while the rear colonnade behaves for winds directions within the ranges [90°, 90°+ $\theta_{F_C}$], and [-90° - $\theta_{F_C}$, -90°].

The complementarity of the acoustic attenuation ranges shows an apparent geometric impossibility of obtaining simultaneous angular protection the lateral and frontal walls of the cell for the same incidence angle, although the acoustic pressure on the frontal surfaces is characterized by lower weight with respect to the lateral ones, according to the ratio between the cell transverse and longitudinal dimensions.

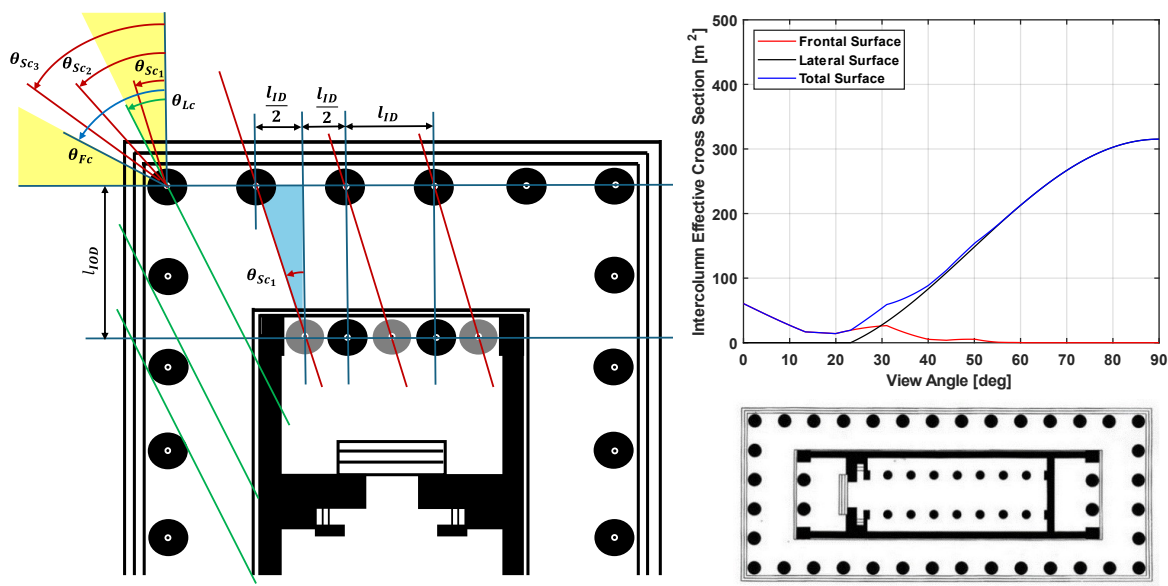

**Figure 8** Hera II Temple in Paestum: vibroacoustic geometric model (top left); intercolumn effective cross section in the view angle range for the view angle [90°- $\theta_{F_C}$, 90°] (top right); scheme of the temple (bottom left); pictures of the temple (bottom right).

A more in-depth acoustic analysis shows, however, that the pronaos and the opisthodomos, semi-open spaces longitudinally bordering the cell longitudinally (Figure1), are actually very effective acoustic attenuators. In fact, the pronaos can be geometrically interpreted as a volume limited by two pairs of parallel walls, the first one consisting of closed (or partially closed) walls parallel to the lateral colonnade, the second one consisting of a closed wall (cell front wall) and of a virtual wall, delimited and partially closed by the internal colonnade, parallel to the external colonnade. As will be clear later, the external attenuating action is carried out by means of a geometric structural mechanism characterized by a synergic action of the two frontal colonnades, external and internal, in fact independent of the specific structural and architectural implementation of their frontal parts.

In principle, an acoustic attenuator structurally similar to the lateral colonnade one would require the virtual closure of the internal colonnade, not achievable for angles, $\theta$, with $\theta < \theta_{F_C}$. Nevertheless, the vibroacoustic analysis shows that there are incidence angles at which the internal colonnade is closed by the acoustic shadows of the external columns, easily evaluated by simple trigonometric relations due to the periodicity of both colonnades.

For the sake of clarity, we will limit ourselves to describing the operating principle of this construction element only for the pronaos of the temple of Hera II (Neptune Temple) in Paestum (Salerno, Italy), although the reasoning is perfectly applicable to all the temples in the sample, despite the construction differences of the front and rear frontal parts (Fig.8). The Hera II's pronaos consists of three closed lateral walls and a semi-open wall limited by only two columns (internal colonnade).

Geometrically, the critical angles, $\theta_{Sc_i}$, for which the internal colonnade is virtually closed (optimal acoustic attenuator) can be expressed, in a completely general way, by the trigonometric relation (Fig. 8)

$$\theta_{Sc_i} = arctg\left[\left(i + \frac{1}{2}\right) \cdot \frac{l_{ID}}{l_{IOD}}\right] \qquad (i = 0, n-1) \qquad (10)$$

where $l_{IOD}$ is the distance between the external and internal colonnade and $n$ is the number of columns, that defines also the number possible of shadow closures. Conversely, the colonnade is open (no shadows between columns) for the angles, $\theta_{Fo_i}$, given by

$$\theta_{So_i} = arctg\left(i \cdot \frac{l_{ID}}{l_{IOD}}\right) \qquad (i = 0, n) \qquad (11)$$

where $n$ is the number of internal columns, while it is closed (virtually closed space between columns). This innovative acoustic-geometric approach shows that there are incidence angles, $\theta_{Sc_i}$, for which the closure condition of the internal colonnade is within the closure range of the external lateral colonnade. The intercolumn effective cross section for the temple of Hera II ($n = 3$), also reported in Figure 8, shows that its best acoustic attenuation (< 20 m²) to winds is in the range [13°-21°].

A direct angular extension of the intercolumn effective cross section representation leads to the definition of a synthetic diagram showing the effective surfaces of the cell exposed to the incident winds, as function of their incidence angle, $\theta$ (Figure 9a). This diagram clearly shows the angular ranges for which the cell appears much more protected from winds, and therefore, from external forcing of its walls.

These results lead to the formulation of the immaterial hypothesis that the temple orientation (angular difference from the east-west direction) could have been defined by the need of optimizing their acoustic attenuation with respect to acoustic forcing of the main and prevailing periodic winds related to the orography of the place. In Figure 9b, a polar diagram showing one-year statistics of wind directions and intensity is shown for the Paestum place.

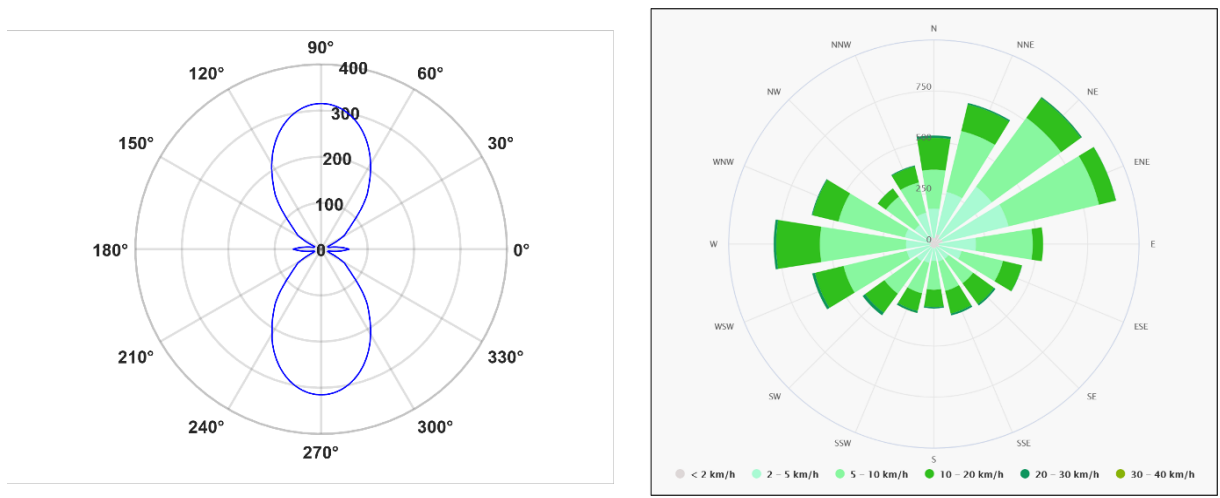

**Figure 9.** (a) Polar acoustic attenuation diagram for the Hera II Temple in Paestum expressed in squared meter (left); (b) statistics of wind direction and intensity of a year period in Paestum (right). The wind rose was retrieved from the web page https://www.meteoblue.com/it/tempo/historyclimate/climatemodelled/pesto_italia_3171709 under Creative Commons license (BY-NC) [accessed on: 22 Feb 2025].

Although limited in time and not referred to the temple construction period, the diagram shows that globally the temple appears to be well oriented, although the few degrees from east-west orientation appears to be only an optimization choice. Nevertheless, the lack of original archaeo-meteorological data could not technically prevent a direct demonstration, since the main winds directions do not appear to have largely changed since then (Murray, 1987). What has been affirmed can be found and confirmed also for the other Doric temples of the sample.

Nevertheless, and this is the reason why the temple of Hera II and the Archaeological Park of Paestum were chosen for such a demonstration even if it has a general validity, the use of available historical data relating to the site coupled with a new reading of the orientation of the temples within the archaeological parks using satellite images, provides sufficient clues to validate the motivations.

**3.6 Application to the Archaeological Park of Paestum**

The archaeological park of Paestum represents an important element in favour of this immaterial vibroacoustic hypothesis of temples' orientation, since it includes the entire city of Paestum, located on the plain of the Sele river, occupying a large area delimited by walls. Inside the walls there are three Doric temples in an excellent state of conservation, one located in the north-east part of the city (Temple of Athena), the other two (Temples of Hera I and Hera II) located in the South-East part.

An aerial view of the city of Capaccio, including also the area of the Archaeological Park of Paestum (Figure10). The aerial view, elaborated with the open source QGIS software, was collected in year 2012, with a 5 cm pixel resolution, and made freely accessible through the National Geoportal of the Italian Ministry of the Environment in the form of a WMS layer (http://wms.pcn.minambiente.it/ogc?map=/ms_ogc/WMS_v1.3/raster/ortofoto_colore_12.map).

This picture clearly shows that the urban part of Paestum archaeological site (buildings and streets), oriented according to a yellow grid superimposed to the aerial view, is oriented differently from the temples, that are oriented according to the red grid represented in the picture. In fact, the streets of the city are oriented almost perfectly in the east-west and north-south directions with orthogonal crossroads, an urban structure typical of classical Greek urban planning, also favoured by the fact that the city is located on a plain. The Doric temples, instead, spread on a very large surface and close to the walls and not classically positioned within a single sacred area, are parallel to each other despite the different construction periods and different position within the walls, oriented at about 4° in the East-South-East direction.

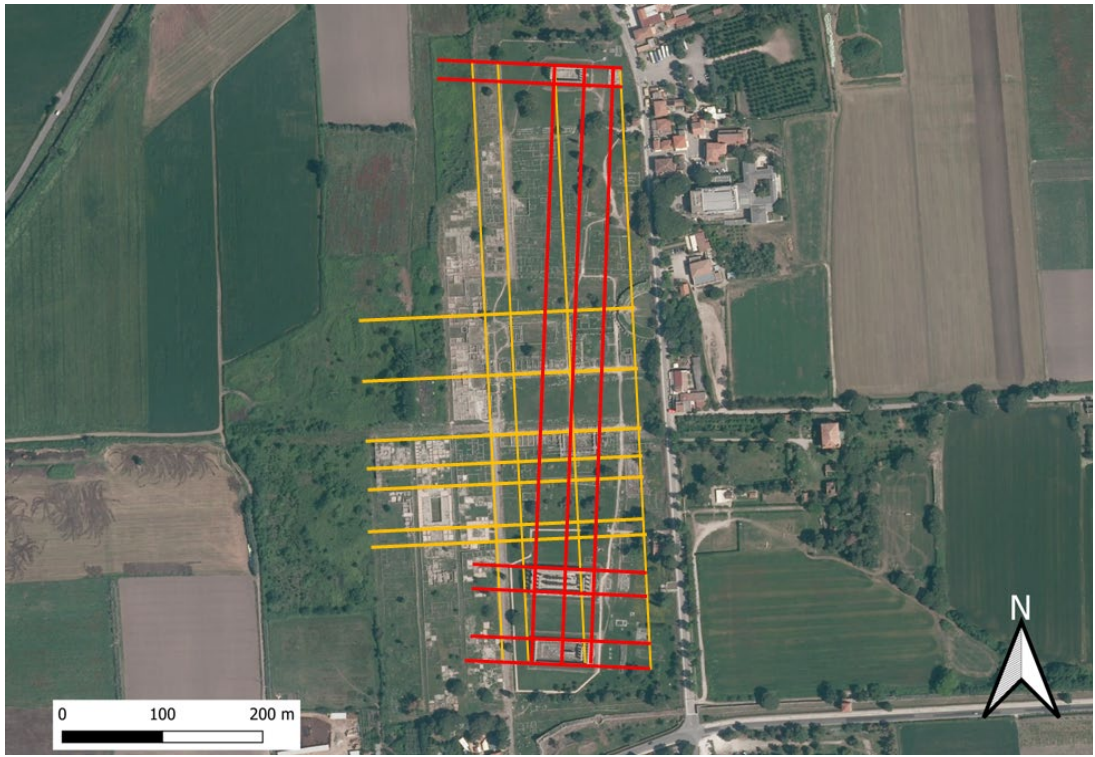

**Figure 10.** Aerial view of the Archaeological Park of Paestum (Paestum, Salerno, Italy). The aerial image was elaborated superimposing two grids: the yellow grid represents the orientation of the buildings; the red grid represents the orientation of the temples. In particular, the bottom part of the red grid embeds the Hera-I and Hera-II temples.

The quality of our hypothesis is confirmed not only by the orientation and parallelism among the three temples visible in Figure 10 (red grid), but also by the preservation of the same orientation of the two recently discovered Doric temples, located on the western border of the city of Paestum (near the walls), even if far from the other ones. demonstrating that the orientation itself is not a coincidence, but a precise orientation related to the Doric temple construction in Paestum, thus confirming that the relevance of the acoustic parameters in the Doric temple construction, part of the immaterial cultural heritage.

## 4. Conclusions

This study, starting from the reanalysis of correlations among geometrical data, developed a vibroacoustic model to interpret the design of Doric temples as a synergy of material and immaterial heritage, emphasizing its integrated structural and functional role beyond aesthetics. Through a multidisciplinary approach, combining archaeological, metrological, and physical analyses, the study demonstrated how Greek architects applied their techne (η τέχνη) to optimize structural resilience and acoustic insulation.

The results highlight the role of lateral colonnades and the peribolos as acoustic attenuators, minimizing wind forces on the cell walls, while the pronaos, opisthodomos, and frontal colonnades function as directional acoustic barriers. These findings suggest that slight deviations from an East-West orientation were deliberate, optimizing acoustic decoupling based on prevailing winds, as evidenced in the Archaeological Park of Paestum.

This interpretation reframes Doric temples as manifestations of advanced design principles deeply as part of Greeks intangible cultural heritage and traditional ecological knowledge, that allowed to combine the architectural design to the local environmental conditions.

Beyond the implications for architectural history, this archaeo-metrology study demonstrates the potential of integrating multidisciplinary methodologies to reveal functional aspects of ancient structures. Future research will refine this model further advancing the understanding of historic architectures as achievements shaped by environmental awareness and the practical knowledge of ancient architects.